\documentclass[a4paper,11pt]{article}
\usepackage[utf8]{inputenc}
\usepackage[english]{babel}

\usepackage{mathtools}
\usepackage{amsmath}
\usepackage{mathtools}

\usepackage[new]{old-arrows}
\usepackage{tikz-cd}

\usepackage{makecell}

\usepackage{amsthm}
\usepackage{amsfonts}
\usepackage[section]{placeins}

\usepackage[font={small,it}]{caption}

\usepackage{float}
\usepackage{hyperref}

\usepackage{tikz}
\usetikzlibrary{matrix,arrows,decorations}
\newtheorem{theorem}{Theorem}[section]

\theoremstyle{remark}
\newtheorem{remark}[theorem]{Remark}
\theoremstyle{definition}
\newtheorem{definition}[theorem]{Definition}

\title{Dynamic sensitivities and Initial Margin via Chebyshev Tensors}

\author{Mariano Zeron\footnote{m.zeron@mocaxintelligence.com}   \\ Ignacio Ruiz\footnote{i.ruiz@mocaxintelligence.com}}

\begin{document}
\maketitle

\begin{abstract}

This paper presents how to use Chebyshev Tensors to compute dynamic sensitivities of financial instruments within a Monte Carlo simulation. Dynamic sensitivities are then used to compute Dynamic Initial Margin as defined by ISDA (SIMM). The technique is benchmarked against the computation of dynamic sensitivities obtained by using pricing functions like the ones found in risk engines. We obtain high accuracy and computational gains for FX swaps and Spread Options.
\end{abstract}

\section{Introduction}\label{sec: Intro}


Sensitivities of portfolios are typically computed every day within banks. Currently, they are used for daily P$\&$L risk management and hedging purposed, as well as for VaR calculations and the associated regulatory capital.


Financial institutions compute forward balance sheet valuations inside Monte Carlo simulations. Examples of these are XVA and IMM capital simulations. However, to our knowledge, no financial institution computes forward sensitivities inside Monte Carlo simulations. Doing such a calculation would give the institution insights into the future \emph{risk} profile, alongside the future balance sheet valuation profiles. The benefits of this would be multi-fold. For example, it would increase the understanding of --- and, hence, capability to manage --- future expected and tail-event hedging needs, future VaR management and subsequent market-risk VaR capital pricing inside KVA, sound pricing of funding cost of future margin, MVA, etc. From these applications, MVA is the one that seems to have gained some popularity in recent years due to the introduction of mandatory margining between financial institutions.

Indeed, one of the consequences of the $2008$ financial crisis has been a worldwide push for strong collateralisation of OTC derivatives. The following table shows the amounts of Variation Margin (VM) and Initial Margin (IM).\footnote{Latest publicly available information at the time of this article going to press. Current numbers should be substantially higher.}

\begin{table}[H]
\centering
\begin{tabular}{|l|l|l|l|}
\hline
 & Cleared ($\$$bn) & Uncleared ($\$$bn) & total ($\$$bn)\\
\hline
Variational Margin & $260.8$ & $870.4$	& $1,131.2$\\
Initial Margin & $173.4$ & $107.1$ & $280.5$\\
Total & $434.2$ & $977.5$ & $1,411.7$ \\
\hline
\end{tabular}
\caption{Cleared and Uncleared Variation and Initial Margin up to September $2017$}
\label{tab: Margin amounts in market}
\end{table}

Out of the two margins, IM should show the highest growth rate in the coming years, potentially going beyond the trillion dollar mark.

Initial margin requirements translate into funding cost, liquidity risk and, potentially, a capital benefit. These requirements are essential in the management of these costs/benefits and risks, both present and future. This requires simulating Initial Margin inside a Monte Carlo (MC) simulation. We call simulated Initial Margin, \emph{Dynamic Initial Margin} (DIM).

Specific uses of DIM include trade pricing (MVA), regulatory capital (IMM and CVA-FRTB), risk management (hedging and tail risk), stress testing and most likely, accounting MVA. Therefore, sound models for DIM will be central for financial institutions; they would particularly shine during a collateral liquidity crisis in which, for example, forward IM simulations done in the past accurately model potential future crisis scenarios. This would allow the institution to be ready for a potential present crisis.

To simplify IM reconciliation between counterparties, the industry has adopted the Standard Initial Margin Model (SIMM) for inter-bank IM posting. This is a model based on the sensitivities of the portfolio to specific risk factors, weighed appropriately by parameters calibrated during periods of stress (see \cite{ISDA SIMM}).

The computational cost of computing DIM using the pricing functions in risk engines is substantial. Assuming an average of $10$-$50$ sensitivities per trade (\cite{ISDA SIMM}), and a typical Monte Carlo simulation with $1,000,000$ nodes, the cost is of the order of $O(10^7)$. This is prohibitively high in practice.

As function approximators, Chebyshev Tensors enjoy strong convergence properties. Once built, they are evaluated very efficiently. Chebyshev Tensors have already been shown to accelerate a wide range of risk calculations (see \cite{MZ IR disclosure}, \cite{MZ IR FRTB}, \cite{GlauParamOptPric}, \cite{Glau_pachon}). In this paper, we use them to compute sensitivities within a Monte Carlo engine (dynamic sensitivities). These sensitivities are then used to compute dynamic SIMM. We show substantial computational reductions (up to $97.5\%$) compared to the benchmark approach, while keeping very high levels of accuracy both at an averaged and tail-event level.

The  paper is organised as follows. Section \ref{sec: cheb tensors} introduces Chebyshev Tensors and the theory that supports the use of these objects in all sorts of risk calculations. Section \ref{sec: Sensitivities with Cheb} presents how to use Chebyshev Tensors to compute dynamic sensitivities. Section \ref{sec: Results} presents the results obtained from the simulation of dynamic sensitivities and dynamic SIMM using Chebyshev Tensors on a FX swap and an European Spread Option. In Section \ref{sec: Results analysis} the accuracy and speed of these two methodologies are compared to the benchmark methodology; the benchmark being sensitivities obtained with original pricing functions at each node of the simulation. The paper closes with a conclusion in Section \ref{sec: Conclusion}.

\section{Chebyshev Tensors}\label{sec: cheb tensors}

Chebyshev Tensors and Chebyshev interpolants lie at the heart of the techniques used to compute the results shown in Section \ref{sec: Results}. This Section covers their main definitions and mathematical properties. For further details we refer the reader to \cite{TrefethenTextbook} and \cite{MoCaXChebUltra}.

\subsection{Chebyshev points and tensors}\label{subsec: cheb points}

Polynomial interpolation enjoyed a bad reputation for a good part of the $20$-th century. This is partly --- among other results --- because interpolation on equidistant points is ill conditioned (Runge phenomenon, \cite{Runge}) and because there is no interpolation scheme that works for continuous functions (\cite{Faber}). Several textbooks on the subject of function approximation warn against them (Appendix in \cite{TrefethenTextbook}). However, what is often missed is that interpolation on carefully selected points can yield optimal approximation properties when applied on the correct class of functions.

\begin{definition}\label{def: cheb pts 1d}
The Chebyshev points associated with the natural number $n$ are the real part of the points
\end{definition}

\begin{equation*}
x_j = \mathrm{Re}(z_j) = \frac{1}{2}(z_j+z_j^{-1} ),\ \ \ \ \ 0\leq j \leq n.
\end{equation*}

Equivalently,

\begin{equation*}
x_j = \mathrm{cos} \Big( \frac{j\pi}{n} \Big), \ \ \ \ \ \ 0\leq j \leq n.  
\end{equation*}

These points are the result of projecting equidistant points on the upper half of the unitary circle onto the real line.

\begin{figure}[H]
\centering
\includegraphics[scale=0.5]{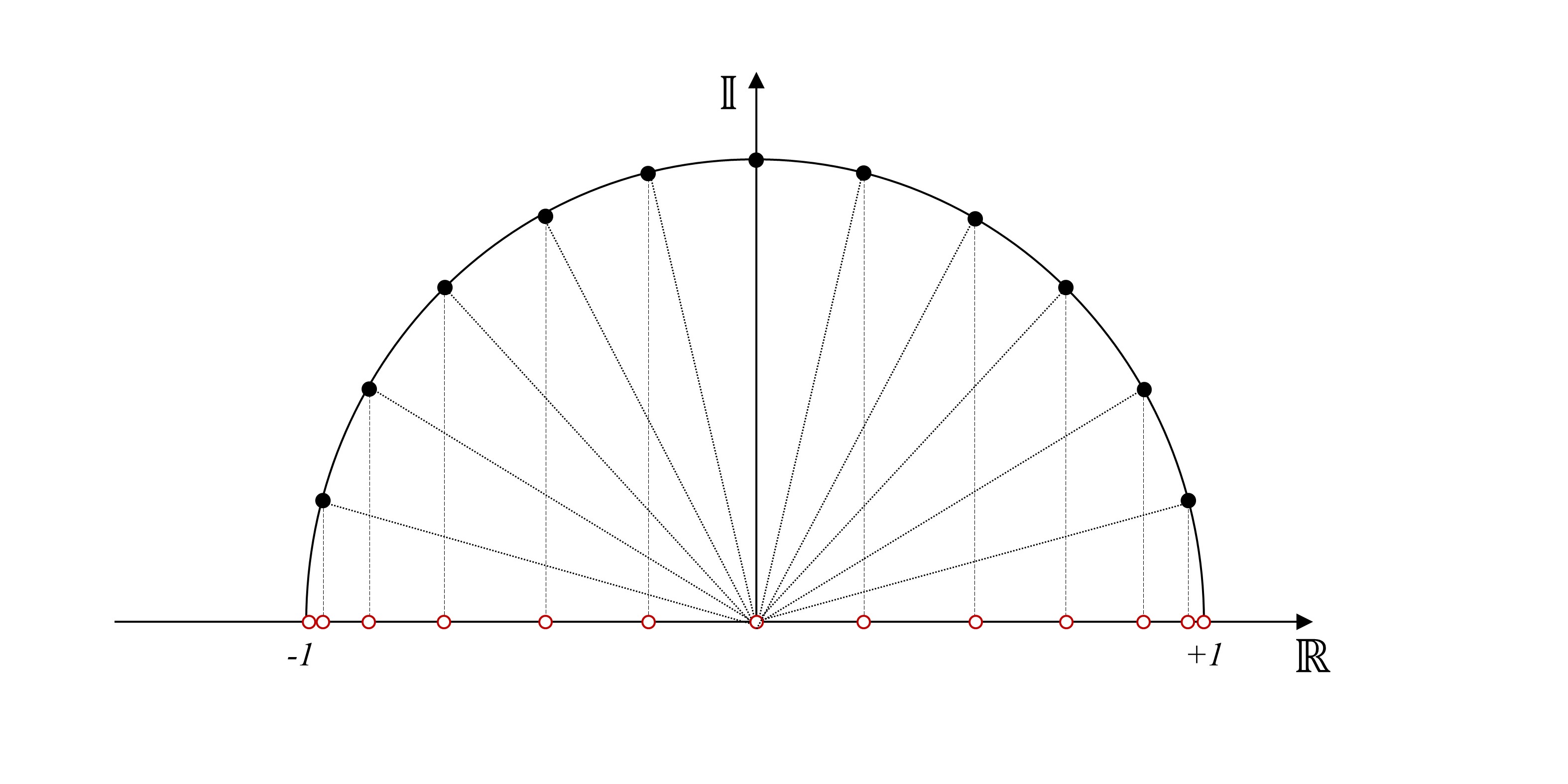}
\caption{Chebyshev points in dimension one.}
\end{figure}

\begin{figure}[H]
\centering
\includegraphics[scale=1]{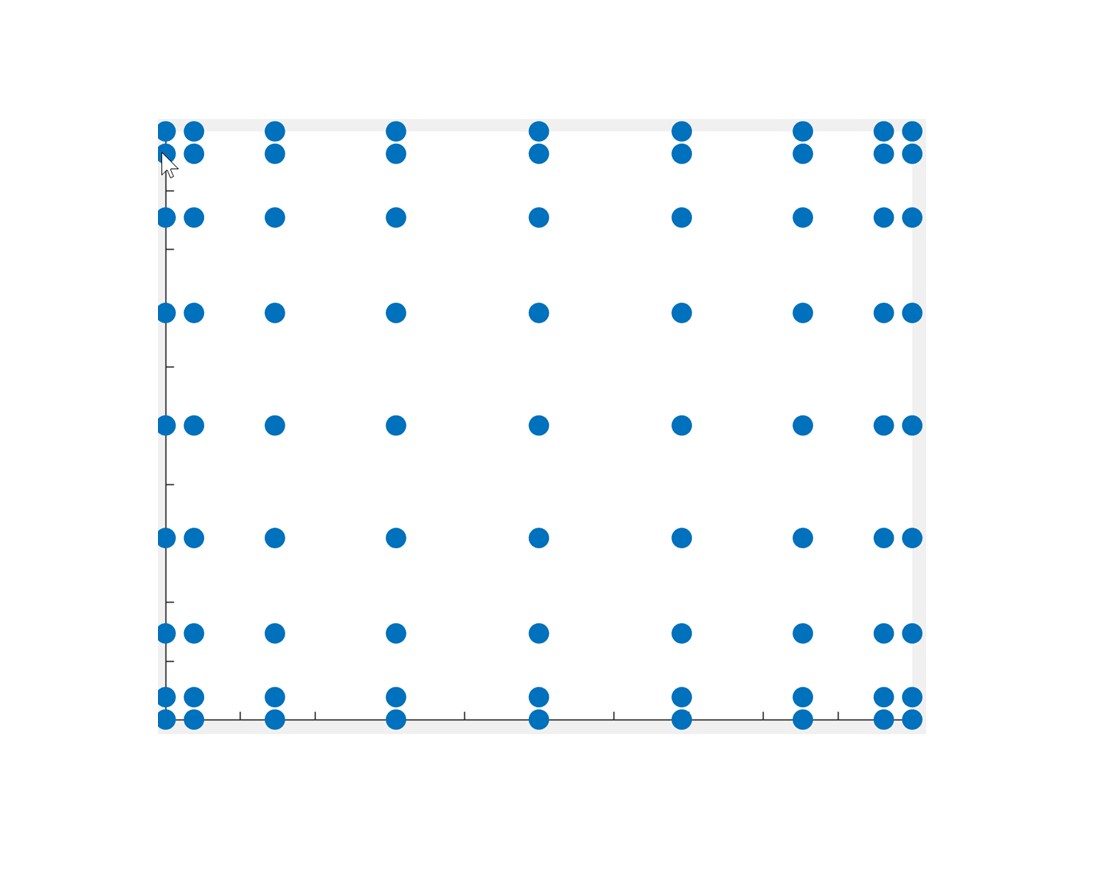}
\caption{Chebyshev points in dimension two.}
\label{fig: cheb pts 2D}
\end{figure}

The extension to higher dimensions is done by taking the Cartesian product of one-dimensional Chebyshev points. Figure \ref{fig: cheb pts 2D} shows an example of a Chebyshev grid in dimension $2$.

Definition \ref{def: cheb pts 1d} is given for an interval $[-1,1]$. This can be extended to any interval $[a,b]$ by appropriately mapping $[-1,1]$ to $[a,b]$. Any results for functions on $[-1,1]$ are valid for more general domains $[a,b]$. The equivalent also applies in higher dimensions.

A \emph{tensor} consists of a set of points $x_0, \ldots, x_n$ in Euclidean space along with a set of associated real values $v_0, \ldots, v_n$. Tensors are closely related to polynomial interpolants. Given a tensor with points $x_0, \ldots, x_n$ and values, $v_0, \ldots, v_n$, there is a unique polynomial $p_n$ of order at most $n$ that interpolates the values $v_0, \ldots, v_n$ at the points $x_0, \ldots, x_n$. When $x_0, \ldots, x_n$ are Chebyshev points, we have a \emph{Chebyshev Tensor} and its corresponding \emph{Chebyshev Interpolant}.

\subsection{Convergence properties}\label{subsec: convergence props cheb}

Chebyshev Tensors have unique convergence properties. Under mild smoothness conditions, convergence is guaranteed. If the function $f$ is Lipschitz continuous, convergence is uniform. If $f$ is smooth, convergence is polynomial. The strongest form of convergence is obtained for analytic functions.\footnote{We remind the reader that a function $f$ is analytic if for all $x$ in the domain of $f$, the Taylor expansion at $x$ converges to $f(x)$.} This is expressed in Theorem \ref{thm: exponential multidim}, which says that very few points are needed to get a high degree of accuracy when the function is analytic (see \cite{GlauParamOptPric} for details).

\begin{theorem}\label{thm: exponential multidim}
Let $f$ be a $d$-dimensional analytic function defined on $[-1,1]^d$. Consider its analytical continuation to a generalised Bernstein ellipse $E_p$, where it satisfies $\|f\|_{\infty}  \leq M$, for some $M$. Then, there exists a constant $C>0$, such that

\begin{equation*}
\|f-p_n \|_{\infty}\leq C\rho^{-m} 
\end{equation*} 

\noindent where $\rho=min_{(1\leq i\leq d)} \rho_i$, and $m=min_{(1\leq i\leq d)} m_i$. The collection of values $\rho_i$ define the radius of the generalised Bernstein ellipse $E_p$, and $m_i$ is the size of the Chebyshev grid for dimension $i$.
\end{theorem}

\subsection{Pricing functions and smoothness}\label{subsec: pricing functions analytic}

There are a couple of comments to be made regarding functions in finance (pricing functions, sensitivity functions, etc.) and how well they lend themselves to be approximated by Chebyshev Tensors.


Pricing functions, outside isolated points, are differentiable and often analytic. Not only is there growing evidence of this (for example \cite{GlauParamOptPric}), but for the most part, practitioners assume so, at least implicitly; the use of Taylor expansions to locally approximate pricing functions is an acknowledgement of this.


The points where pricing functions are not differentiable usually come in the form of payment dates, barrier, strikes, etc. An important thing to note about these points is that it is possible to locate them within the domain of the function; often, they are defined by the trade itself. To prevent these points from affecting the convergence properties of Chebyshev Tensors, one splits the domain of approximation along these points. By doing so, one is left with a collection of sub-domains free of singularities, over which Chebyshev Tensor enjoy the properties mentioned in Theorem \ref{thm: exponential multidim}.


Singular points can also be the result of structured payoffs. For example, taking the maximum between continuation and exercise value in an American option introduces a singularity. However, the continuation function is free of this type of singularity and carries nearly all the computational cost. Hence, Chebyshev Tensors are built for this function obtaining the expected benefits.


It is important to note that convergence rates of Chebyshev Tensors are determined by the smoothness of the function (see Section \ref{subsec: convergence props cheb}). How non-linear the trade is does not impact its smoothness. Linear products will likely need less Chebyshev points than non-linear ones, but the type of convergence does not change.


Another case to consider is that of pricing functions that rely on simulations. For these, the values are themselves approximations of the true value. What the authors have observed empirically is that Chebyshev Tensors approximate the function up to the level of accuracy provided by the implementation of the pricing function. For example, if prices are obtained through Monte Carlo simulations and the simulations come with a noise of $1e^{-3}$, then the Chebyshev Tensor will reach this level of accuracy exponentially, but will remain within this noise regardless of added Chebyshev points.




\subsection{Tensor Extension Algorithm}\label{subsec: tensor extension algo}

Tensors are normally affected by the curse of dimensionality. However, this is not an issue in the computation of dynamic sensitivities. This is the direct result of how Chebyshev Tensors are applied in this context (Section \ref{sec: Sensitivities with Cheb}), and the Tensor Extension Algorithm that we shall now discuss.

\subsubsection{Tensors in TT format}\label{sec: tt format}

The Tensor Extension Algorithms presented in \cite{glau_completion} works with tensors expressed in TT format. These are tensors that admit a representation which is memory efficient, allowing the use of high dimensional tensors in practical settings where otherwise it would be impossible. Here we only present a summary of \cite{glau_completion}. For more details we refer to \cite{glau_completion} and \cite{Completion_Steinlechner}.




%
Consider vectors $C_i$ of matrices with pre-specified rank $r_{i-1}\times r_i$. That is, for all $j$, such that $1\leq j\leq n_i$, $C_i(j)$ is a matrix of rank $r_{i-1}\times r_i$. If the ranks for all $C_i$ are given by $(r_0, r_1,\ldots, r_d)$, where $r_0 = r_d = 1$, then we can define the following tensor $\mathcal{X}$ as follows

\begin{equation}\label{eq: tensor TT rank r}
\mathcal{X}(i_1, \ldots, i_d) = C_1(i_1)\cdots C_d(i_d).
\end{equation}

Notice that the product of matrices $C_1(i_1)\cdots C_d(i_d)$ makes sense given the way the ranks have been defined and that all the information needed to recover the tensor is contained in the vectors $C_i$. The memory cost of $\mathcal{X}$ when storing the values at every grid point is $\mathcal{O}(n^d)$; when expressed as in Equation \ref{eq: tensor TT rank r}, only $\mathcal{O}(dnr^2)$ --- what was exponential growth in terms of dimension is now linear.

Apart from the potentially huge memory cost reductions, tensors in TT format have another important property for the applications of interest in this paper. When defined on Chebyshev points, these tensors can be very efficiently evaluated. The evaluation of these tensors is given through the inner product of tensors defined as follows

\begin{equation}\label{eq: tensor inner product}
\langle \mathcal{X}_1, \mathcal{X}_2 \rangle = \sum_{i_1 = 1}^{n_1} \cdots \sum_{i_d = 1}^{n_d}\mathcal{X}_1(i_1, \ldots, i_d)\mathcal{X}_2(i_1, \ldots, i_d),
\end{equation}

The details of how the inner product is used to evaluate Chebyshev Tensors is beyond the scope of this paper. Details can be found in \cite{glau_completion}. The key takeaway is that the evaluation is very efficient as it amounts to little more than the product of matrices of low rank; routines which are highly efficient in most numerical packages these days.

\subsubsection{Approximation with Tensors in TT format}\label{subsec: approx with tensors in TT}

Say we want to build a tensor $\mathcal{T}$ for a function $f$ evaluating $f$ on the whole set of grid points and storing them is problematic or impossible due to the sheer amount of data. However, if we can approximate $\mathcal{T}$ with a tensor $\mathcal{X}$ in TT format --- which, as explained in Section \ref{sec: tt format}, are much cheaper to store --- then $\mathcal{X}$ may be used as proxy for $f$. This is what the Tensor Extension Algorithms presented in \cite{glau_completion} do.

There are three algorithms that constitute the Tensor Extension Algorithms as presented in \cite{glau_completion}: the \emph{Completion Algorithm}, the \emph{Rank Adaptive Algorithm}, and the \emph{Sample Adaptive Algorithm}. Each builds on top of the previous. We briefly describe the main ideas behind each of them.

The Completion Algorithm is designed to find tensors $\mathcal{X}$ in TT format with fixed rank. Let $\mathcal{F}_{r}$ be the space of TT tensors with rank $r = (r_0, \ldots, r_d)$. An important property of $\mathcal{F}_{r}$ is that it has the structure of a smooth Riemannian manifold. The algorithm makes use of this as one can explore such spaces in an efficient manner by using algorithms such as the Riemannian Conjugate Gradient.

At each iteration of the algorithm, a candidate $\mathcal{X}$ is considered and compared to the tensor $\mathcal{T}$ on a fraction of its grid. Note that if we had $\mathcal{T}$ on all its grid we would not need to find $\mathcal{X}$. The problem is that the dimension of $\mathcal{T}$ is such that evaluating it on all its grid is problematic if not impossible. Therefore, we evaluate $\mathcal{T}$ on a sub-grid $\mathcal{K}$ and compare every candidate $\mathcal{X}$ to $\mathcal{T}$ on $\mathcal{K}$. If the error achieved is deemed low enough, the algorithm stops. In summary, the Completion Algorithm uses the information of a fraction of the grid $\mathcal{K}$ and completes this information to the whole of a tensor in TT format.

Just like in a Machine Learning model, the Completion Algorithm splits the sub-grid $\mathcal{K}$ into two subsets. One of the subsets is used to find the tensor $\mathcal{X}$; this can be considered the training set. The other subset is used for testing.

The Completion Algorithm searches spaces of fixed rank. The \emph{Rank Adaptive Algorithm} runs the Completion Algorithm, increasing the rank of the space each time if no suitable $\mathcal{X}$ is found. Intuitively, the greater the rank, the bigger the space and the higher the chances of finding a suitable $\mathcal{X}$.

Both the Completion and the Rank Adaptive Algorithm work with a fixed sub-grid $\mathcal{K}$. The purpose of the \emph{Sample Adaptive Algorithm} is to increase the size of $\mathcal{K}$ when the Rank Adaptive Algorithm does not yield a good result. The algorithm stops when either a suitable $\mathcal{X}$ is found or when a pre-established limit on the number of grid points evaluated is reached.

It is important to note that there is no guarantee these algorithms will find a suitable $\mathcal{X}$ to approximate $\mathcal{T}$. However, if the function $f$ to approximate is well behaved, one normally expects these algorithms to give good results. The results presented in section \ref{sec: Results} can be taken as empirical evidence that for some pricing functions in finance, these algorithms can indeed find suitable tensors $\mathcal{X}$ to work with. Further evidence of this is presented in \cite{glau_completion}.

\section{Computing dynamic sensitivities with Chebyshev Tensors}\label{sec: Sensitivities with Cheb}

Consider a Risk Factor Evolution Model (RFEM) that generates risk factors in a Monte Carlo simulation. Take, for example, the Hull-White one-factor model 

\begin{equation}\label{eq: hull white equation}
dr_t = a (b - r_t)dt + \sigma dW_t
\end{equation}

It has parameters $\theta = (a, b, \sigma)$ and only one stochastic variable $W_t$. Define the \emph{model space} as the space spanned by the short rate $r$. Let the dimension of the model space be $k$. In this example $k$ = 1. For a two-factor HW model, $k = 2$. For most models, the dimension is the same as the number of stochastic variables. In the context of Monte Carlo simulations for XVA or IMM, $k$ tends to be small. 

Once the parameters $\theta$ of the RFEM have been calibrated, they remain fixed throughout the simulation. At every node of the simulation, the model space variables (short rate $r$, for example) fully determine the market risk factors (e.g. a full swap rate curve) for that node. The latter may include interest rates, spreads, volatilities, etcetera. We call the space of market risk factors the \emph{market space}. The latter typically has high dimension; sometimes in the hundreds (e.g. collections of swap rates and implied volatility surfaces). Denote the dimension of the market space by $n$.

Denote the function that generates market risk factors from model space variables by $g$

\begingroup
\Large
\begin{equation}\label{eq: g function model space}
\begin{tikzcd}
\stackrel{  \textbf{ Model Space } } { \mathbb{R}^k}  \rar{g}  &  \stackrel{  \textbf{ Market Space  } } { \mathbb{R}^n}  
\end{tikzcd}
\end{equation} 
\endgroup

In the case of the Hull-White one factor model, $g$ is given by

\begin{equation*}
\Big(S(t, T_1), \ldots, S(t, T_n)\Big) = \Big(A(t, T_1)e^{B(t, T_1)r(t)},\ldots, A(t, T_n)e^{B(t, T_n)r(t)}\Big),
\end{equation*}

\noindent where $r(t)$ is the short rate, $T_i$ a time point ahead of $t$, and where $A$ and $B$ only depend on the parameters $\theta$ and today's yield curve.

Functions like $g$ are often analytic and lend themselves very well to be approximated with Chebyshev Tensors. 

The following example shows how Chebyshev Tensors can be built to compute sensitivities within a Monte Carlo simulation. A Foreign Exchange Swap is used as an example; this is the trade type for which results are presented in Section \ref{sec: Results analysis}. Whatever is said equally applies to any other trade type.

\subsection*{Example}

Let the pricing function of a FX Swap be $f$. The market risk factors sensitivities needed are with respect to swap rates (for two different currencies) and the exchange rate. Say there are $n$ risk factors in total.

Each sensitivity --- as a function of market risk factors --- has dimension $n$. The value of $n$ tends to be such that building a tensor for it is problematic because of the curse of dimensionality. Therefore, the dimension of the problem must be reduced. The following approach, which makes use of the RFEM, is the one we propose. 

Consider a single time point within the Monte Carlo simulation. The sensitivity of $f$ to each market risk factor must be computed at each scenario. As an example, let the $i$-th swap rate be $s_i$. Consider the following function $\varphi$

\begingroup
\Large
\begin{equation*}
\begin{tikzcd}
\mathbb{R}^k\rar{\widetilde{g}}\arrow[black, bend right]{rr}[black,swap]{\varphi}  & \mathbb{R}^n \rar{S_i}  & \mathbb{R}
\end{tikzcd}
\end{equation*}
\endgroup

\noindent where $S_i$ denotes the partial derivative of $f$ with respect to $s_i$ 
\begin{equation*}
S_i = \frac{\partial f}{\partial s_i}.
\end{equation*}

Note that $\widetilde{g}$ is the result of putting together the parametrisations $g$ (such as the ones in Equation \ref{eq: g function model space}) of the RFEMs used to diffuse the market risk factors that correspond to the trade. In the case of the FX Swap there are five models: each currency has two yield curves, one for forward projections and one for discounting. Each curve was simulated using a two-factor Gaussian model as RFEM. Additionally, the FX rate was simulated with a one-factor log-normal Brownian motion model. This gives a total of $k = 9$ dimensions.

The $k$-dimensional function $\varphi$ is the result of composing two analytic functions. Therefore, by Theorem \ref{thm: exponential multidim} Chebyshev Tensors will converge to them quasi-exponentially as the number of grid points in each dimension increases. Given its definition, the function $\varphi$, gives the value of the partial derivative of $f$ with respect to $s_i$ at each node of the simulation. This means that the Chebyshev Tensor directly approximates the sensitivity.

To build a Chebyshev Tensor for $\varphi$ do the following. Take the minimum and maximum value of each of the model space variables at the time point in question of the Monte Carlo simulation. These values determine the hyper-rectangle to which $\varphi$ is restricted. Notice, the hyper-rectangle just mentioned is contained in $\mathbb{R}^k$. Next, build a Chebyshev grid on this hyper-rectangle.

Once the hyper-rectangle in $\mathbb{R}^k$ is ready, one must decide how to build the Chebyshev Tensor: either directly, by evaluating each grid point, or by using one of the Tensor Extension Algorithms presented in Section \ref{subsec: tensor extension algo}. If the dimension $k$ of $\varphi$ is between $1$ and $4$, one can build a Chebyshev Tensor directly. For example, if the trade is an Interest Rate Swap, where each yield curve can be modelled with $1$ or $2$ factor models, yielding a tensors of $2$ or $4$ dimensions. If $k$ is greater than $4$, one ought to consider the use of the Tensor Extension Algorithms. As explained in Section \ref{subsec: tensor extension algo}, to build a Chebyshev Tensor this way, one evaluates a fraction of the grid points and the algorithm returns a Chebyshev Tensor in TT format that is a proxy to the tensor one would have obtained if the whole of the Chebyshev grid had been evaluated.
\bigskip

\begin{remark}\label{rmk: suing time to mat as var}
The description given so far corresponds to Chebyshev Tensors built for each market risk factor at each time point of the Monte Carlo simulation. However, one can also consider the time dimension in the construction of the Chebyshev Tensor. In this case, the domain of $\varphi$ increases in dimension by one. This may be a suitable thing to do in some cases but not all. For example, if the trade has payments as it matures, one should consider the discontinuities that arise, as explained in Section \ref{subsec: pricing functions analytic}. If there are many discontinuities, then there will be many tensors to build and perhaps the time point approach is more direct. If there are no discontinuities, instead of building a tensor per risk factor per time point, one only builds one per risk factor, considerably increasing computational gains. This is the case, for example, of the Spread Option for which we show results in Section \ref{sec: Results}.
\end{remark}

\begin{remark}\label{rmk: technique for price}
The focus of this article is the dynamic simulation of sensitives and subsequent DIM. However, the same technique can be applied to pricing functions $f$ instead of sensitivity functions $S_i$. This generates Chebyshev Tensors that efficiently compute portfolio price simulations for the standard XVA or IMM Monte Carlo simulations. In this case, the computational benefits are down to the same reasons: Chebyshev Tensors approximate pricing functions with high levels of accuracy; the approximation is obtained by calling the potentially slow-to-compute pricing function $f$ a relatively small number of times; the evaluation of Chebyshev Tensors is highly efficient, especially when compared to sophisticated pricing functions.
\end{remark}

\section{Results}\label{sec: Results}


Dynamic sensitivities were computed using Chebyshev Tensors --- built using the Tensor Extension Algorithms presented in Section \ref{subsec: tensor extension algo} --- in two Monte Carlo simulations. One for a FX Swap; the other for a European Spread Option. The objective was to use the sensitivities to compute IM at each scenario of the Monte Carlo simulation and DIM profiles.\footnote{Note that the version of IM computed is SIMM, the one proposed by ISDA, and by now a standard in the industry for uncleared derivative transactions.}




The accuracy of the technique is measured with respect to the benchmark. The latter is obtained by computing dynamic sensitivities using the original pricing functions, either through finite differences in a ``brute force" fashion --- as was the case for the FX Swap --- or directly as a result of the pricing routine --- as was the case for the Spread Option based on Monte Carlo simulations. The time taken using Chebyshev Tensors in each case is measured and compared to the time taken to perform the benchmark calculation.


Note that DIM profiles should be computed at netting set level. Our tests only considered a couple of trades, each constituting its own netting set. The technique employed naturally extends to a netting set with multiple trades by applying it trade by trade.


The tests were done in MATLAB, on a standard laptop with i7 cores. Calculations were parallelised whenever possible, in particular for the benchmark calculations, given their high computational cost.

\subsection{FX Swap}\label{subsec: fx swap results}


The FX Swap was between USD and EUR, at-the-money and had $5$ years to maturity. The pricing routine used was the one implemented in the swapbyzero function in MATLAB, which is analytic. Sensitivities were obtained by finite difference using this pricing routine.


The Monte Carlo simulation used consisted of $10,000$ paths and $11$ time points, covering the full lifespan of the trade.


The ISDA risk factors affecting the FX Swap consist of a collection of swap rates in two currencies (USD and EUR) and the exchange rate. Each currency has two yield curves; one for forwarding and one for discounting. The yield curves were diffused using a $2$-factor Gaussian model. The exchange rate was diffused using Geometric Brownian Motion.


FX Swaps have payment dates. These create discontinuities in the pricing function along the time dimension. Therefore, as explained in Section \ref{sec: Sensitivities with Cheb}, Chebyshev Tensors were built per sensitivity and per time point in the simulation.\footnote{It was decided not to sub-divide the domain along the time dimension in this example. Therefore, time to maturity is a constant parameter for each Chebyshev Tensor.}


Given the RFEM used and the market risk factors involved, the tensors had dimension $9$. The domain over which they were built was determined by the range of the model space risk factors diffused at each time point of the simulation. Four points per dimension were used to build the Chebyshev grid. This gives a grid with $262,144$ points. This tensor would take a long time to build directly. Therefore, the Sample Adaptive Algorithm presented in Section \ref{subsec: approx with tensors in TT} was used.


The Sample Adaptive Algorithm started with $300$ points for training and $50$ for testing. That is, $350$ function evaluations. As the algorithm increases the number of points and, hence, required function evaluations, if it deems it necessary, a maximum number of evaluations needs to be set. This was fixed at $1,000$. For most risk factors and time points, $300$ points were more than enough to reach an accuracy at both training and testing of $5e^{-3}$. In some cases the number of points needed was $500$. The time taken for the algorithm to run varied, but never beyond $1$ minute; in some cases just seconds. The average time taken for the Chebyshev Tensors to evaluate all scenarios on a time point of the simulation was of $15.5$ seconds. The corresponding time for the benchmark method was $1,100$ seconds (see Section \ref{sec: Results analysis} for details on computational savings).

\begin{figure}[H]
\centering
\includegraphics[scale=0.4]{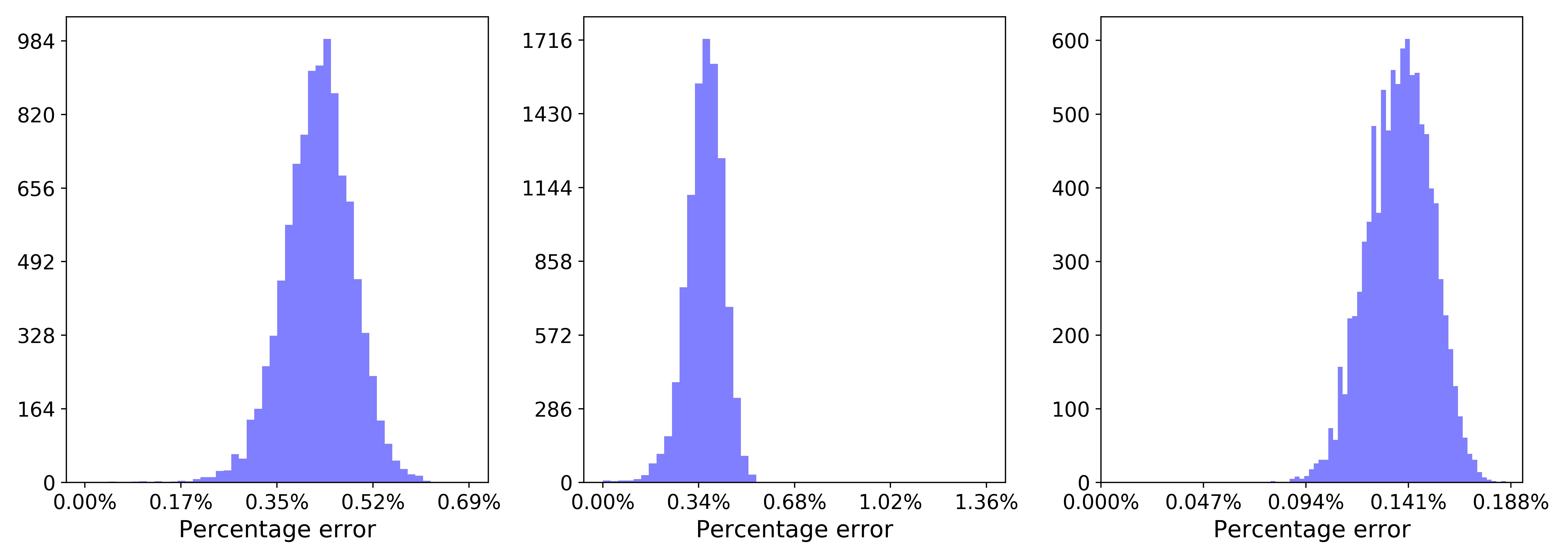}
\caption{Percentage relative errors of Chebyshev Tensors for the first swap rate of the USD forwarding yield curve. Histograms correspond from left to right, to the second, sixth and eleventh time point in the simulation.}
\label{fig: sens errors FX rf 9}
\end{figure}
\begin{figure}[H]
\centering
\includegraphics[scale=0.4]{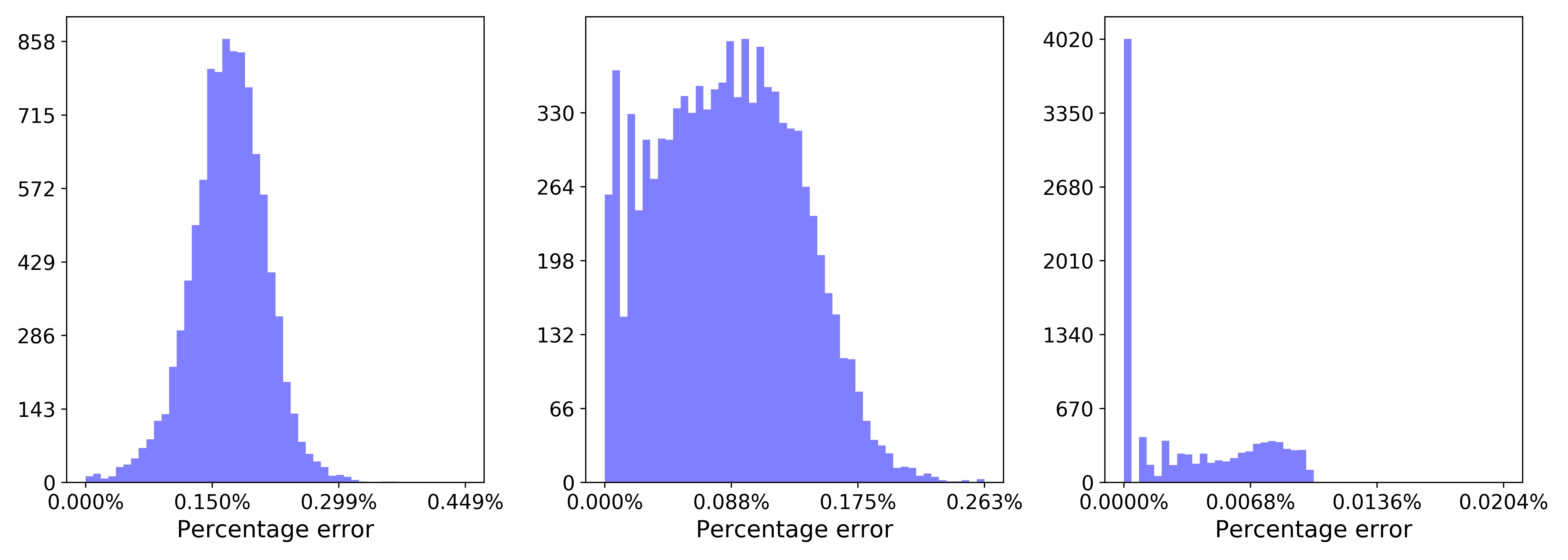}
\caption{Percentage relative errors of Chebyshev Tensors for the USD/EUR exchange rate. Histograms correspond from left to right, to the second, sixth and eleventh time point in the simulation.}
\label{fig: sens errors FX rf 33}
\end{figure}

The following Figures show relative error distributions of the Chebyshev Tensors built to compute dynamic sensitivites for two risk factors at three different time points of the simulation. Figure \ref{fig: sens errors FX rf 9} shows the errors for the first swap rate of the USD forwarding yield curve. Figure \ref{fig: sens errors FX rf 33} shows the errors for the exchange rate. Note that the vast majority of the errors are well under $1\%$. In fact, for all the risk factors considered, the maximum error was always below $1.5\%$ (see Table \ref{tab: relative errors fx swap}).

\begin{figure}[H]
\centering
    \includegraphics[scale=0.35]{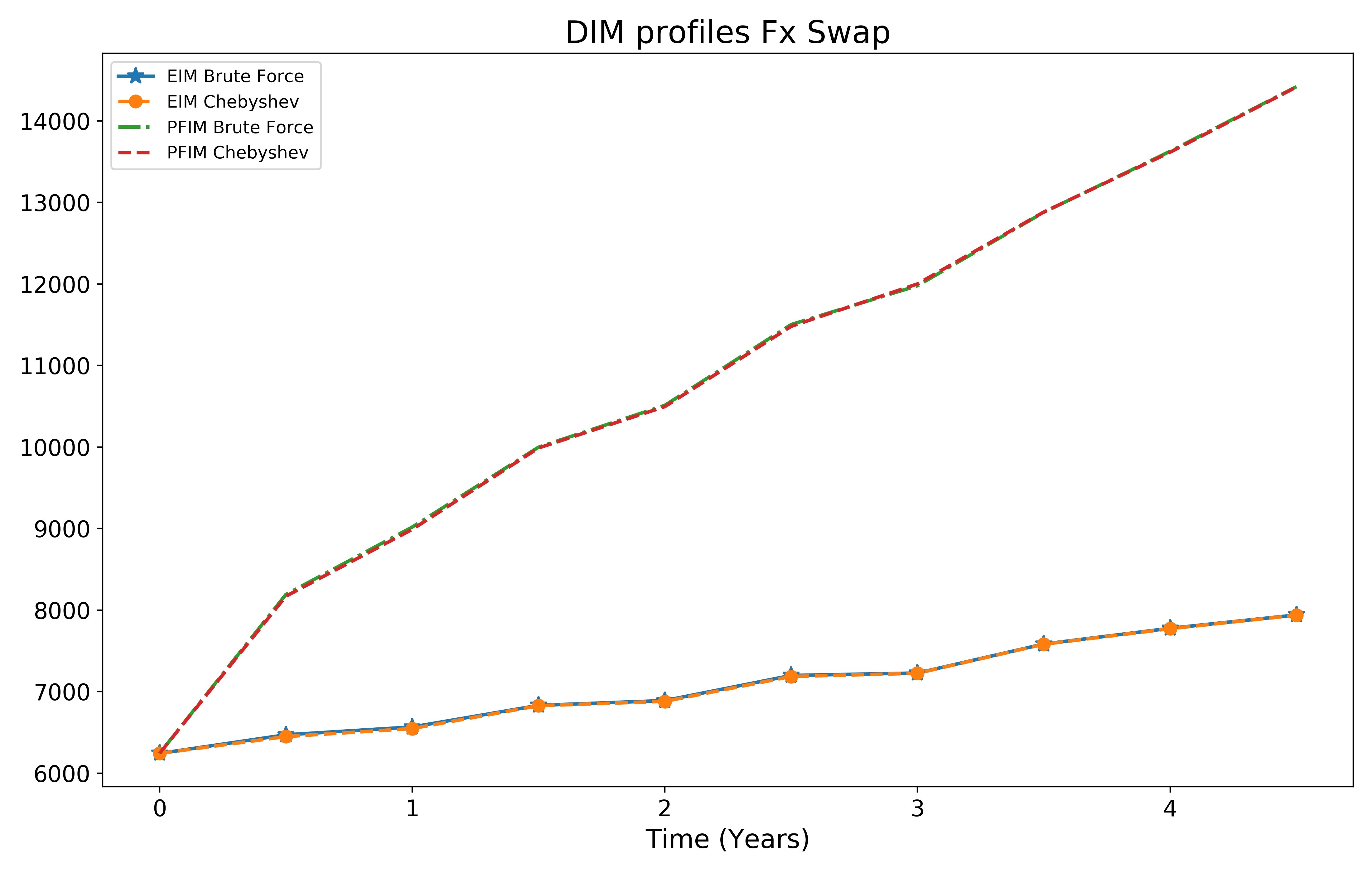}
\caption{DIM profiles --- expectation and $95\%$ quantiles --- for FX Swap obtained with the benchmark and with Chebyshev Tensors.}
\label{fig: DIM profiles fx swap}
\end{figure}

Define EIM as the Expected Initial Margin and PFIM as the Potential Future Initial Margin at $95\%$ confidence level, at each future time point. Figure \ref{fig: DIM profiles fx swap} shows EIM and PFIM profiles obtained with the benchmark and with Chebyshev Tensors. As can be seen in Figure \ref{fig: DIM profiles fx swap} and Table \ref{tab: relative errors fx swap}, the errors are all below $0.34\%$.

\begin{table}[H]
\centering
\renewcommand{\arraystretch}{1.35} 
{\footnotesize
\begin{tabular}{|c|c|c|c|}
\hline
\textbf{FX Swap} & \textbf{ISDA sensitivities} & \thead{\textbf{EIM}} & \thead{\textbf{PFIM}}  
 \\\hline
Maximum relative error & $1.5\%$ & $0.34\%$ & $0.32\%$ \\
\hline
\end{tabular}
}

\caption{Maximum relative percentage error for market sensitivities, EIM and PFIM ($95\%$ quantile) profiles for the FX Swap.}
\label{tab: relative errors fx swap}
\end{table}

\subsection{European Spread Option}


The chosen European Spread option had one year to maturity. The pricing routine used was Monte Carlo-based. That is, we have a nested Monte Carlo simulation. The outer simulation, for sensitivities and DIM, consisted of $10,000$ paths. The inner simulation, for the local derivative pricing routine, consisted of $1,000$ paths.

The inner simulation was run with antithetic variates, and the average evaluation time was of $0.5$ seconds per sensitivity. Given the evaluation time and the size of the Monte Carlo simulation for the computation of dynamic sensitivities, a full benchmark calculation took, even with parallelisation, up to ten hours for each risk factor.

Given the number of times the pricing routine was called for the benchmark calculations, the number of paths within the pricing routine had to be kept reasonably low for practical reasons. With $1,000$ paths, the $95\%$ quantile of the noise of spot sensitivities (i.e. the delta) was measured at $7.4\%$. Reducing the noise by half would require (roughly) increasing the number of paths by an order of magnitude; from $1,000$ to $10,000$. This would take the benchmark computation of dynamic sensitivities, for spot, from $10$ hours to several days. Given the lack of computational power available, the decision was made to stick to $1,000$ paths in the pricing function and work with the fact that Chebyshev Tensors would be accurate to at most the noise level. 

The situation was worse for the sensitivities of the remaining risk factors (i.e. non-spot risk factors). For Vega sensitivities, the $95\%$ quantile of the noise was measured at $32.7\%$; for swap rates it was measured at $11.23\%$. Correcting this would increase computational times considerably. Therefore, only delta dynamic sensitivities are presented.

\begin{figure}[H]
\centering
\includegraphics[scale=0.4]{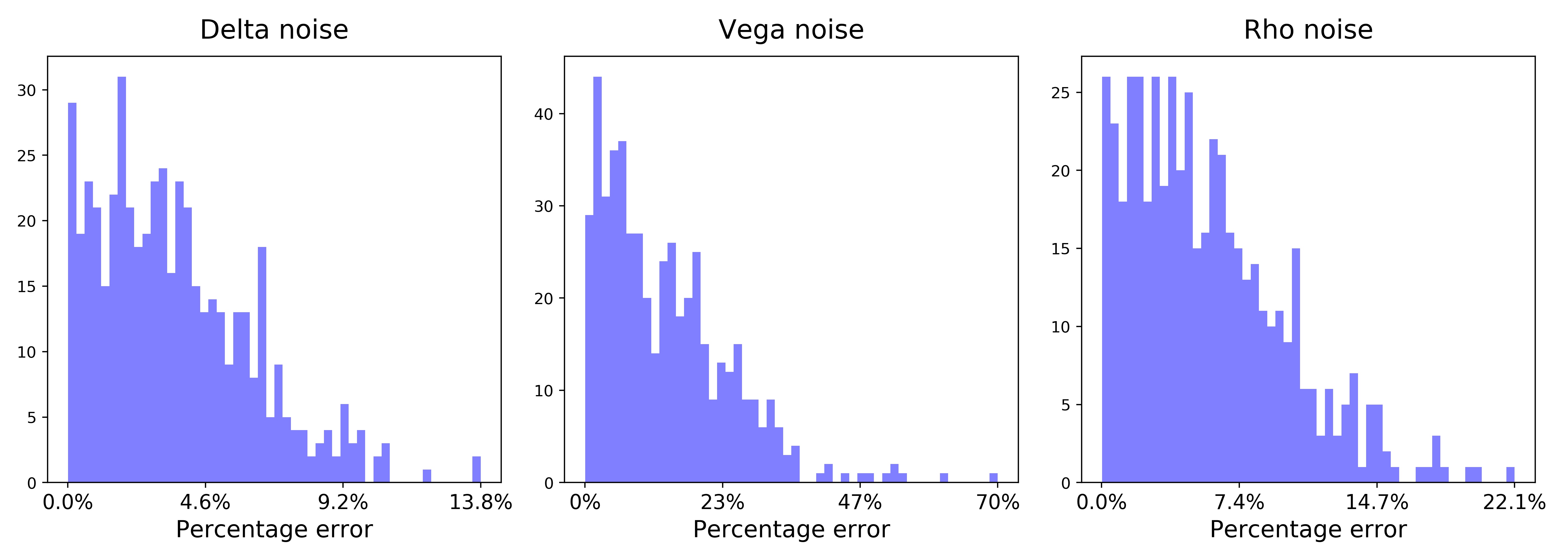}
\caption{Noise distribution for the Monte Carlo-based Spread option pricing function. Histograms correspond from left to right, to the noise for Delta, Vega and Rho.}
\label{fig: noise pricer spread}
\end{figure}

Ideally, we would like to report results for an American-style option. However, due to the time it would take to produce benchmark results, this is not possible.\footnote{This is due to two reasons. Firstly, the American-style pricing function takes around five times longer than its European equivalent for the same number of paths. Secondly, the numerical noise of the American-style pricer is far greater than its European equivalent for the same number of paths. As a result, the number of paths in the American-style pricer needed to be increased considerably in order to obtain a reasonable numerical noise.} As a result, we report results on an European-style option.

However, as was highlighted in Section \ref{subsec: pricing functions analytic}, the type of payoff does not hinder the properties of the Chebyshev method proposed in this paper. It is therefore expected that, with enough computational power, similar results would be obtained for the American version of this trade.

The Monte Carlo simulation for the computation of dynamic sensitivities consisted of $10,000$ paths and $11$ time points in the future, covering the full lifespan of the trade.


No discontinuities are present along the time dimension for this trade. Therefore, the Chebyshev Tensors built included time to maturity as a variable (see Remark \ref{rmk: suing time to mat as var}). This means that only one tensor per market risk factor was built.


The Spread Option takes two spot underlyings, two volatilities (one for each underlying), and a yield curve. The underlyings were diffused using the Heston model, which diffuses both the spot and the volatility stochastically. The yield curve was diffused using a Hull-White $1$-factor model. Given the dimensions of the RFEMs just mentioned and that time to maturity was considered a variable, the tensors built had $6$ dimensions: $5$ for the model space variables diffusing market risk factors, and $1$ for the time to maturity.


The tensor domains were built by considering the minimum and maximum values attained by the model space variables diffused in the Monte Carlo simulation, along with the time to maturity of the trade. Six points per dimension were chosen for the grid. This gives a total of $46,656$ grid points. Once again, the Sample Adaptive Algorithm was used to obtain a Chebyshev Tensor in TT format that approximates the sensitivities needed. The algorithm started by evaluating $12,000$ random grid points; $10,000$ were used for training and $2,000$ for testing. After a $2$-$3$ minutes, the algorithm typically reached an error of $5e^{-3}$; comparable to the noise level of the pricing function. Once built, the Chebyshev Tensors took an average of $11.1$ seconds to evaluate all scenarios on each time point. The corresponding time for the benchmark approach was measured at $5,000$ seconds. Section \ref{sec: Results analysis} discusses computational savings.





\begin{figure}[H]
\centering
\includegraphics[scale=0.4]{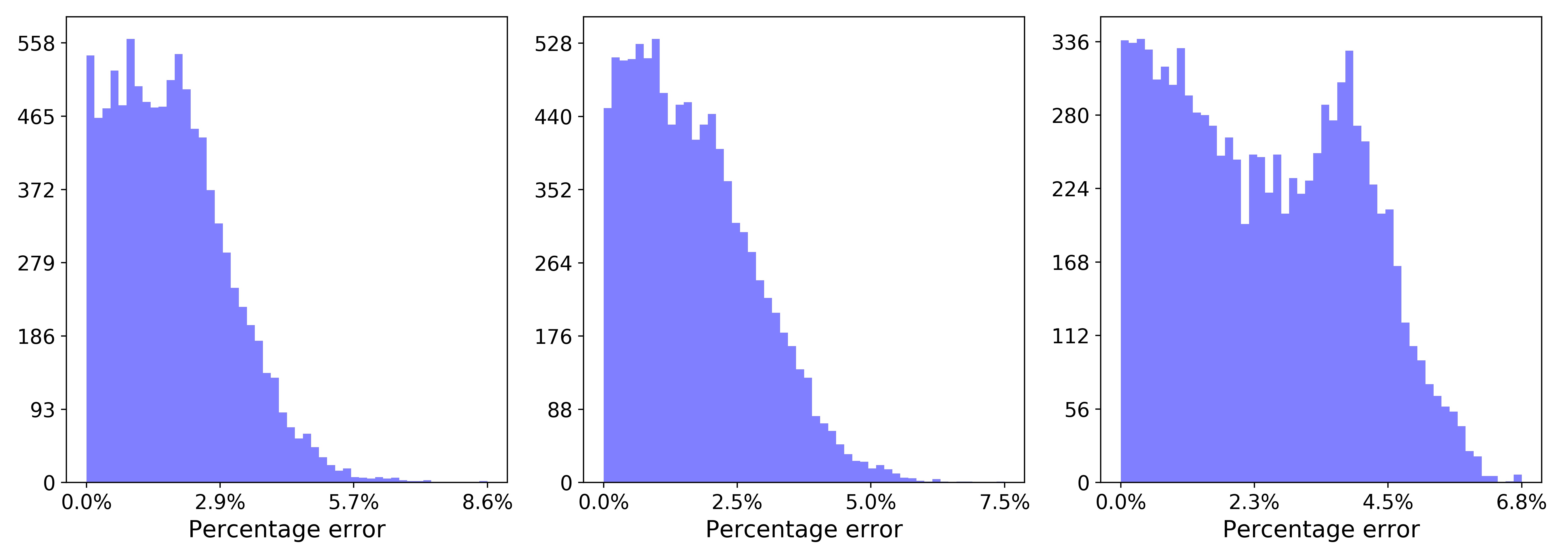}
\caption{Percentage relative errors of Chebyshev Tensors for the first spot. Histograms correspond from left to right, to the second, sixth and eleventh time point in the simulation.}
\label{fig: sens errors spread rf 1}
\end{figure}

Figure \ref{fig: sens errors spread rf 1} shows relative error distributions of the Chebyshev Tensor built to compute dynamic sensitivities for one of the spot underlyings in the Spread Option, at three different time points of the simulation. Note that most errors measured are within the noise of the Delta function presented in Figure \ref{fig: noise pricer spread}.

\begin{figure}[H]
\centering
\includegraphics[scale=0.35]{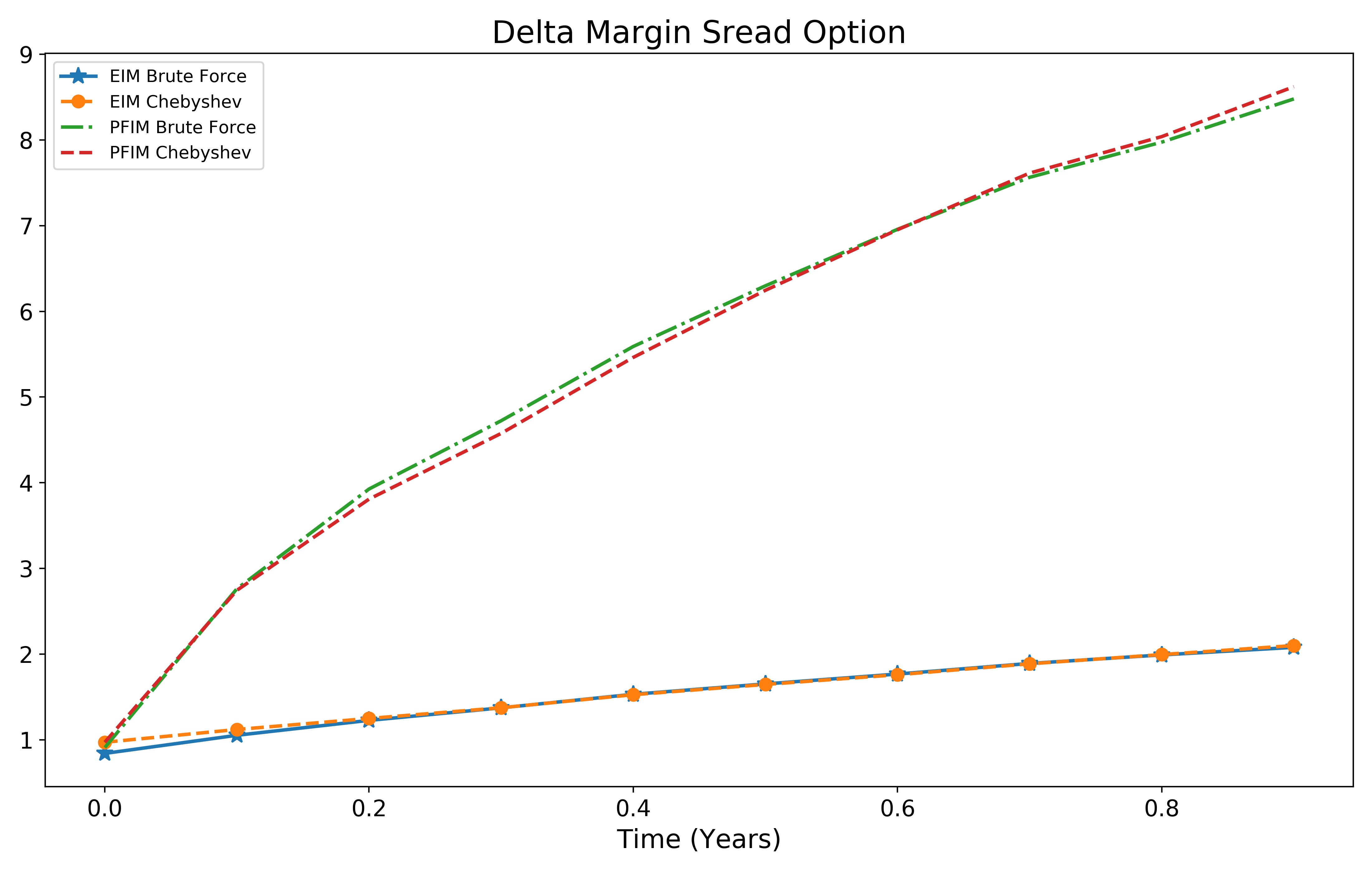}
\caption{Equity Delta Margin profiles --- at expectation and $95\%$ quantiles --- for the European Spread Option obtained with the benchmark and with Chebyshev Tensors.}
\label{fig: DIM profiles spread option}
\end{figure}

Figure \ref{fig: DIM profiles spread option} shows equity Delta Margin profiles, as defined by SIMM, at expectation level (EIM) and $95\%$ quantiles (PFIM) obtained with the benchmark and with Chebyshev Tensors.

The maximum error for market sensitivities was $10.7\%$. However, the vast majority of the errors were below $5\%$, as the ones presented in Figure \ref{fig: sens errors spread rf 1}. The maximum errors at the level of Delta Margin profiles were $8.1\%$ and $3.1\%$, for EIM and PFIM, respectively (Table \ref{tab: relative errors Spread}). Given the noise presented in Figure \ref{fig: noise pricer spread}, the main source of error in the EIM and PFIM comes from the noise intrinsic in the approximated function itself.

\begin{table}[H]
\centering
\renewcommand{\arraystretch}{1.35} 
{\footnotesize
\begin{tabular}{|c|c|c|c|}
\hline
\textbf{Spread Option} & \textbf{Market sensitivities} & \thead{\textbf{EIM}} & \thead{\textbf{PFIM}}  
 \\\hline
Maximum relative error & $10.7\%$ & $8.1\%$ & $3.1\%$ \\
\hline
\end{tabular}
}

\caption{Maximum relative percentage error for Market sensitivities, EIM and PFIM Delta Margin profiles for the Spread Option. EIM profile corresponds to Expected IM, while PFE to $95\%$ quantile.}
\label{tab: relative errors Spread}
\end{table}

\section{Discussion of Results}\label{sec: Results analysis}






For the FX Swap, the level of accuracy for dynamic sensitivities was high all the way through the simulation and for all risk factors. The highest relative error for sensitivities measured at $1.5\%$. However, for almost all risk factors, the maximum relative error was below $1\%$. This translated into small errors at the level of DIM profiles where both EIM and PFIM had maximum relative errors less than $0.35\%$ (see Table \ref{tab: relative errors fx swap}).

These results were expected. The pricing function is analytic and the trade linear; few Chebyshev points are needed to approximate the sensitivities to a high degree of accuracy. This is not only reflected in the high accuracy values, but also in the short training times of the Sample Adaptive Algorithm (see Table \ref{tab: Tensor extension details and comp savings})

The levels of accuracy are such that Chebyshev Tensors should price MVA stably and provide sound measures of MVA sensitivities, allowing for sound hedging. Moreover, the dynamic sensitivities simulations can be used for hedging simulations. The fast calculations allow for fast scenario analysis. It can also be used inside portfolio optimisation routines in order to minimise future IM funding costs.

The computational savings for the FX Swap are summarised in Table \ref{tab: Tensor extension details and comp savings}. Computing benchmark sensitivities, for a single risk factor, on a single time point, requires $20,000$ calls to the pricing function. The Sample Adaptive Algorithm needed between $300$ and $500$ calls. The training of the algorithm took seconds in most cases. The evaluation of the Chebyshev Tensors on the $10,000$ scenarios took an average of $15.5$ seconds. Compared to the cost of the benchmark calculation, for each risk factor and each time point --- measured at $1,100$ seconds --- the training and evaluation times reported for Chebyshev Tensors are negligible. The computational savings are therefore around $97.5\%$.

It must be appreciated that the tests were carried out using MATLAB on a standard laptop computer. The computational times could be far smaller in a bank's optimised pricing library. However, the relative gains the technique brings should be the same.

\begin{table}[H]
\centering
\renewcommand{\arraystretch}{1.35} 
{\footnotesize
\begin{tabular}{|c|c|c|c|c|c|}
\hline
 & \thead{\textbf{Benchmark} \\ \textbf{evaluations}} & \thead{\textbf{Train $\&$ Test} \\ \textbf{evaluations}} & \thead{\textbf{Avg training} \\ \textbf{time}} &  \thead{\textbf{Cheb Tensor} \\ \textbf{eval time}} & \thead{\textbf{Comp} \\ \textbf{savings}} \\
\hline
FX Swap & $20,000$ & $300\sim 500$ &  $<$ minute & $15.5$ secs & $97.5\%$\\
\hline
Spread Option & $110,00$ &	$12,000$   & $2\sim 3$ minutes & $11.1$ secs &  $89\%\sim 98.9\%$\\
\hline
\end{tabular}
}

\caption{Computational savings obtained by using Chebyshev Tensors to compute dynamic sensitivities compared to benchmark method.}
\label{tab: Tensor extension details and comp savings}
\end{table}

The Spread Option uses a Monte Carlo-based pricer. In this case, the accuracy achieved by the single Chebyshev Tensor built to approximate Delta is within the noise reported in Figure \ref{fig: noise pricer spread} (see Figure \ref{fig: sens errors spread rf 1} and Table \ref{tab: relative errors Spread}). This accuracy translates into similar accuracy levels at the level of Delta Margin profiles, as can be seen in Figure \ref{fig: DIM profiles spread option} and Table \ref{tab: relative errors Spread}.

For the Spread Option, computing the benchmark dynamic sensitivities for a single risk factor, within the whole Monte Carlo simulation, required $110,000$ calls to the pricing function ($10,000$ paths, $11$ time points). The Sample Adaptive Algorithm only needed $12,000$. The training of Sample Adaptive Algorithm took between $2$ and $3$ minutes per risk factor. The evaluation of the Chebyshev Tensors took on average, at each time point, $11.1$ seconds. Given the cost of the benchmark sensitivities calculation --- estimated at $83.3$ minutes per risk factor per time point --- the training and evaluation of Chebyshev Tensors is negligible. Therefore, the computational savings are estimated --- given number of time points used --- at $89.1\%$. A typical Monte Carlo simulation consists of $100$ time points. Increasing the number of time points in the simulation does not change the way Chebyshev Tensors are built --- notice time to maturity is included as a variable. Therefore, for a Monte Carlo simulation with $100$ time points, the computational savings would be of $98.9\%$.






\subsection{Pre-trade analysis}


Pre-trade analysis has recently become a much-desired feature of XVA, IMM capital and PFE systems; more generally of CCR metrics. This consists of measuring the impact on CCR metrics of possible incoming trades. Given the time constraints of the business and the time it takes to compute sensitivities within the Monte Carlo simulation (or PVs, depending on the calculation), this is normally not possible with benchmark pricing functions. 

Chebyshev Tensors help accelerate this what-if type of analysis. With $90\%+$ computational savings, in some cases it is possible to build them on the fly for each potential incoming new trade. If not, one should build a Chebyshev Tensor that includes, as part of the variables in its domain of approximation, all those parameters that differentiate instances of the same trade-type. 

For example, say a Spread Option must be incorporated into the netting set. By including strike as part of the Chebyshev Tensor, one builds an object capable of approximating a wide range of Spread Options. As long as the maturity and strike are within the domain of the tensor, this can be used to obtain the sensitivities of all possible Spread Options within the simulation. Given the speed of evaluation for Chebyshev Tensors, this can be done in a short period of time allowing for adequate pre-trade analysis.




\section{Conclusion}\label{sec: Conclusion}

Chebyshev Tensors enjoy remarkable mathematical properties that make them ideal candidates to approximate analytic functions --- such as pricing functions --- to high degrees of accuracy with little computational effort (Section \ref{sec: cheb tensors}). This paper shows how to harness the power of Chebyshev Tensors through the use of Tensor Extension Algorithms (Section \ref{subsec: tensor extension algo}), to compute dynamic sensitivities and with these, Dynamic Initial Margin (dynamic SIMM), to a high degree of accuracy, with low computational cost, and limited implementation effort. 

The technique tested an at-the-money FX Swap, with $5$ years to maturity, and a European Spread Option, with $1$ year to maturity. The benchmark computation consisted of dynamic sensitivities obtained by calling the pricing function --- such as the ones found in Front Office systems --- at each node of the simulation.


In the case of the FX swap, the maximum relative error for dynamic sensitivities was $1.5\%$, while for DIM profiles was $0.34\%$ (Section \ref{subsec: fx swap results}). For the Spread Option, maximum relative error for dynamic deltas was $10.7\%$, mostly due to the numerical noise of the underlying function being approximated. Only Delta Margin was computed in this case; the errors for the corresponding profiles stood at $8.1\%$ and $3.1\%$, which is in-line with the noise presented in the pricing function itself (Figure \ref{fig: noise pricer spread}).


Computational gains stand at $97.5\%$ for the FX swap and between $89\%$ and $98.9\%$ for the Spread Option (Table \ref{tab: Tensor extension details and comp savings}).



There is nothing in the Chebyshev method presented in Section \ref{sec: Sensitivities with Cheb} that prevents it from being applied to a wide range of trade types. As mentioned in Section \ref{subsec: pricing functions analytic}, linearity and payoff type do not hinder the use of the methodology. The key element to consider is the dimension of the tensor to build. In CCR, most of the models used are of a dimension which the Tensor Extension Algorithms in Section \ref{subsec: tensor extension algo} can handle. Problems can appear in cases such as Basket Options with a large number of underlyings modelled independently. For these cases, the algorithms become more costly to run, reducing computational gains. However, for a typical portfolio, we expect the technique to apply for the vast majority of live and newly incoming trades.

The natural extension to this paper consists of running the calculations shown for a wider range of trades, in particular ones with higher dimensions. This is left for a future project.

\pagebreak

\end{document}